\documentstyle[preprint,tighten,aps,psfig,epsf]{revtex}

\newcommand{\be}{\begin{equation}}
\newcommand{\ee}{\end{equation}}
\newcommand{\ba}{\begin{eqnarray}}
\newcommand{\ea}{\end{eqnarray}}

\newcommand{\fr}[2]{\frac{#1}{#2}}
\newcommand{\non}{\nonumber}

\def\vec#1{{\mbox{\boldmath$#1$}}}

\newcommand{\p}{\mbox{$\vec{p}$}}
\newcommand{\q}{\mbox{$\vec{q}$}}

\newcommand{\k}{\mbox{$\vec{k}$}}

\newcommand{\r}{\mbox{$\vec{r}$}}

\newcommand{\lb}{\left (}
\newcommand{\rb}{\right )}
\newcommand{\la}{\left\langle}
\newcommand{\ra}{\right\rangle}

\newcommand{\sS}{\mbox{$\vec{\sigma}\vec{\sigma}'$}}

\begin{document}

\preprint{TTP99-05, hep-ph/9902276}

\title{ ${\cal O}(m \alpha^7 \ln^2 \alpha)$ corrections to
positronium energy levels}

\author{Kirill Melnikov\thanks{
e-mail:  melnikov@particle.physik.uni-karlsruhe.de}}
\address{Institut f\"{u}r Theoretische Teilchenphysik,\\
Universit\"{a}t Karlsruhe,
D--76128 Karlsruhe, Germany}
\author{ Alexander Yelkhovsky \thanks{
e-mail: yelkhovsky@inp.nsk.su }}
\address{ Budker Institute for Nuclear Physics\\
Novosibirsk, 630090, Russia}
\maketitle

\begin{abstract}
We present a calculation of the
${\cal O}(m \alpha^7 \ln^2 \alpha)$
corrections to positronium energy levels. The result is used 
to estimate the current uncertainty in  theoretical
predictions of the positronium spectrum.
\\
\vspace{0.5cm}
\end{abstract}

1. Recently, a calculation of the positronium spectrum with 
${\cal O}(m\alpha^6)$ accuracy was completed \cite{KP,CMY}.
The remaining uncertainty in  theoretical
predictions was estimated in \cite{CMY} using the value
of the ${\cal O}(m \alpha^7 \ln^2 \alpha)$ leading logarithmic corrections 
to positronium energy levels:
\be
\delta E = - \lb \fr{499}{15}+7\sS \rb
\fr{m\alpha^7 \ln^2 \alpha}{32\pi n^3} \delta_{l0}.
\ee
The purpose of this Letter is to 
present a detailed derivation of this result.

In general, an appearance of logarithms of the fine structure 
constant in  QED bound state problems is related to 
the fact that several momentum scales 
are involved in bound state calculations. Contributions that 
are  logarithmic in $\alpha$, usually appear as integrals of the form:
\be
\int \fr{d^3 k}{(2\pi)^3} F(k),
\ee
with $F(k) \sim k^{-3}$ for the values of $k$ such that 
$m\alpha \ll k \ll m$ or
$m\alpha^2 \ll k \ll m\alpha$ \cite{KMYlog}.
Given the inequality $k \ll m$, it is
possible to determine  the leading  logarithmic corrections in
the framework of the nonrelativistic  Quantum Mechanics
by expanding  all perturbations in series of $k/m$ and
by using the time-independent perturbation theory.

In order to find ${\cal O}(m\alpha^7 \ln^2 \alpha)$ corrections, 
we first calculate all effective operators which
deliver ${\cal O}(m\alpha^n \ln \alpha)$
contributions for $n < 7$. Such contributions first appear
for $n=5$ and $n=6$. Then, the ${\cal O}(m\alpha^7\ln^2\alpha)$ 
corrections to energy levels are found by calculating
relativistic corrections to these lower order
operators. Some new operators arising at ${\cal O}(m\alpha^7)$
order should be also taken into account. For completeness  
we consider here the bound state of two different particles 
with masses $m$ and $M$.

2. Let us first  consider  the self energy operator of one particle
in the Coulomb field of the other.
The scattering amplitude\footnote{The notations of
\cite{CMY} are used throughout this paper.}
reads:
\be\label{Ase}
-A_{\rm SE} = - \alpha \int \fr{d^3 k}{(2\pi)^3}
                \fr{4\pi}{2k}
               j_i(\p',\p'-\k)e^{-i\k\r'_e}
               \fr{
               \delta_{ij} - \fr{k_i k_j}{k^2}
               }{ k+H-E }
               j_j(\p-\k,\p)e^{i\k\r_e}.
\ee
The leading order logarithmic contribution is obtained
by  expanding the denominator  $(k+H-E)^{-1}$
in  $(H-E)/k$ and by using the leading non-relativistic
approximation for the currents $j_i$.
Only spin-independent part of the currents should be
considered since the spin-dependent part contains additional
powers of $k$ which destroy the logarithmic integration. One obtains:
\be\label{vseLO}
V_{\rm SE}^{\rm LO} \to
        \int \fr{d^3 k}{(2\pi)^3}
        \fr{2\pi\alpha}{k^3}
        \fr{\p_k e^{-i\k\r_e} }{ m }
        \lb H-E \rb \fr{\p_k e^{i\k\r_e} }{ m },\;\;\;\;\;
\p_k \equiv \p-\k(\p\k)/k^2.
\ee
It is easy to see that the integration over $k$ in the
above equation is already logarithmic. For this reason
we can neglect the non-commutativity of the Hamiltonian
$H$ with the exponent $\exp(i\k\r_e)$. Fusing two exponents
together, we integrate over $k$ cutting the integral
by the reduced particle mass $\mu = mM/(M+m)$
from above and by the particle energy $\sim p^2/\mu$ from below.
Including the self energy of the second particle, we obtain an
effective operator:
\be\label{4}
V_{\rm SE}^{\rm LO}(\p,\r) \to
           - \fr{2\alpha}{3\pi\mu^2}
             \lb 1-2\fr{\mu^2}{mM} \rb
             \p (H-E) \p
             \ln \fr{p^2}{\mu^2},
\ee
Here and below we consider  $\ln (p^2/\mu^2)$ as
commuting with all other quantities.

By averaging this operator over $nS$ states,
one obtains the well-known non-recoil logarithmic contribution to
the Lamb shift of the $nS$ levels:
\be\label{selo}
\delta_{\rm SE}^{\rm LO} E =
           - \fr{8\alpha^2\psi(0)^2}{3\mu^2}
             \lb 1-2\fr{\mu^2}{mM} \rb \ln\alpha.
\ee

In order to obtain ${\cal O}(m\alpha^7\ln^2 \alpha)$ corrections
induced by the operator $V_{\rm SE}^{\rm LO}$ it is necessary to
calculate relativistic corrections to the expectation
value of this operator. There are several sources of such
corrections; below we analyze them.

The simplest one is the relativistic correction to the
currents. It is obtained by the substitution
\be
\vec{j}(\p',\p) \approx \fr{\p'+ \p}{2m}
        \to
        -\fr{p'^2+p^2}{4m^2} \fr{\p'+ \p}{2m}
\ee
performed for one of the currents in Eq.(\ref{vseLO}).
Again, only the orbital part of the currents
contributes. The correction to the energy levels reads:
\ba
\delta_{\rm SE}^{\rm curr} E &=&\la
           \int \fr{d^3k}{(2\pi)^3}
           \fr{4\pi\alpha}{2k^3}
           \left[ H, \fr{\p_k e^{-i\k\r_e} }{ m }
           \right]
           \fr{\p_k e^{i\k\r_e} }{ m } \fr{p^2}{m^2}
           \ra \non \\
     &\to& -\la \int \fr{d^3k}{(2\pi)^3}
           \fr{8\pi\alpha\mu}{3m^4 k^3}
           C(r)[\p,C(r)]\p
           \ra \non \\
     &\to& \fr{16\alpha^4\psi(0)^2}{3\mu^2}
           \lb 1-4\fr{\mu^2}{mM}+2\fr{\mu^4}{m^2M^2}
           \rb
           \ln^2\alpha.
\label{SEcurr}
\ea
At the final stage of the calculation the self energy of the second 
particle was included.

The next-to-next-to-leading order effect of the retardation is described by
the operator
\be\label{vseNNLO}
V_{\rm SE}^{\rm ret} = - \int \fr{d^3 k}{(2\pi)^3}
           \fr{\pi\alpha}{k^5}
           \left[ H, \fr{\p_k e^{-i\k\r_e} }{ m }
           \right]
           \left[ H,
           \left[ H, \fr{\p_k e^{i\k\r_e} }{ m }
           \right]\right]
           + {\rm h.c.}
\ee
Though there is the fifth power of $k$ in the
denominator, the spin-dependent part of the
currents still does not contribute to the double
logarithmic correction. It can be
easily seen, that the commutator
$[H,\vec{\sigma}\times\k \exp(i\k\r_e)]$ is ${\cal
O}(k^2)$ as $k\to 0$,
so that the resulting integral over $k$
for the spin-dependent part of the currents
is non-logarithmic. We therefore calculate the
commutators in Eq. (\ref{vseNNLO}) and
keep there only such terms that are quadratic in $k$.
We obtain:
\ba
\delta_{\rm SE}^{\rm ret} E &=&\la
           \int \fr{d^3k}{(2\pi)^3}
           \fr{4\pi\alpha}{3m^3k^3}
           \lb 3+\fr{\mu}{m} \rb
           C(r)[\p,C(r)]\p \ra
           + (m \leftrightarrow M)
           \non \\
     &\to& -\fr{8\alpha^4\psi(0)^2}{3\mu^2}
           \lb 4-13\fr{\mu^2}{mM}+2\fr{\mu^4}{m^2M^2}
           \rb
           \ln^2\alpha.
\label{SEret}
\ea

Next, we consider  relativistic
corrections to the Coulomb interaction and to the
dispersion law of the particles in the intermediate
state:
\be\label{dC}
                    \int \fr{d^3 k}{(2\pi)^3}
                    \fr{2\pi\alpha}{k^3}
                    \fr{\p_k e^{-i\k\r_e} }{ m }
   \left[ \frac {\p^4}{8m^3}+\frac {\p^4}{8M^3}
   + \fr{\pi\alpha}{2} \lb \fr{1}{m^2} + \fr{1}{M^2}
      \rb \delta(\r) \right]
                    \fr{\p_k e^{i\k\r_e} }{ m }.
\ee
Taking the average value of this operator, one sees
that the double logarithmic contribution is absent.

Accounting for additional magnetic exchange between the
particles requires some care. There exist eight
irreducible diagrams. Only four of them, 
shown in Fig.1, deliver double logarithmic contributions.
One finds, that the double logarithmic
contributions from the diagrams Fig.1a and Fig.1b compensate
each other, while that of Fig.1c and Fig.1d sum up to the
following energy shift:
\ba
\delta_{\rm SE}^{\rm magn} E &=&
           - \fr{\alpha}{2mM}
           \la \int \fr{d^3k}{(2\pi)^3}
           \fr{4\pi\alpha}{3k^3}
           \fr{\p}{ m }
           \fr{1}{r}p^2
           \fr{\p}{ m } \ra
           + (m \leftrightarrow M)
           \non \\
           &\to&
           -\fr{4\mu^2\alpha^4}{3\pi mM}
           \lb 1-2\fr{\mu^2}{mM} \rb
           \la \fr{\ln (\mu r)}{r^3} \ra
           \non \\
           &\to&
           -\fr{8\alpha^4\psi(0)^2}{3mM}
           \lb 1-2\fr{\mu^2}{mM} \rb \ln^2\alpha.
\label{SEBreit}
\ea
Integration over $k$ is performed in the limits
$(\mu r^2)^{-1} < k < \mu$.

Finally, there exists ${\cal O}(\alpha^2)$ correction to the
wave function of the bound state, i.e. the correction due
to the iteration of the Breit Hamiltonian and the LO
operator $V_{\rm SE}^{\rm LO}(\p,\r)$ (cf. Eq.(\ref{4})). 
The calculation of this  correction is described 
in the final part of this Letter,
where all lowest order logarithmic operators are
considered simultaneously.

3. We consider now an exchange 
of a single magnetic photon between the
two particles. The corresponding scattering amplitude is
similar to Eq. (\ref{Ase}):                           
\be\label{vM}
-A_{\rm M} = \alpha \int \fr{d^3 q}{(2\pi)^3}
             \fr{4\pi}{2q}
             J_i(-\p',-\p'-\q)e^{-i\q\r'_p}
             \fr{
             \delta_{ij} - \fr{q_i q_j}{q^2}
             }{ q+H-E }
             j_j(\p-\q,\p)e^{i\q\r_e}
             + {\rm h.c.}
\ee
We used this amplitude in \cite{CMY} when discussed
the retardation effects.
Considering the next-to-leading order 
retardation, one finds the operator
\be\label{vMLO}
V_{\rm M}^{\rm LO} \to - \int \fr{d^3 q}{(2\pi)^3}
                    \fr{4\pi\alpha}{mM q^3}
                     e^{i\q\r}
                    \p_q (H-E) \p_q,
\ee
where again the non-commutativity of $e^{i\q\r}$ and $H$ has
been neglected. Because of the presence of the
$\exp(i\q\r)$, the logarithmic integral over $q$ 
in Eq.(\ref{vMLO}) is cut
at $1/r \sim p$ from above \cite{KMYlog}.
We arrive at the following effective operator:
\be\label{15}
V_{\rm M}^{\rm LO}(\p,\r) = - \fr{2\alpha}{3\pi mM}
                        \p (H-E) \p
                        \ln \fr{p^2}{\mu^2}.
\ee

The sources of double logarithmic
corrections to the single magnetic exchange
are the same as in the case of the  self energy operator. The only
essential difference in two calculations
consists in a change of the upper cut-off.
For this reason we skip a detailed discussion and
present the results of the calculation.
We obtain:
\ba
\delta_{\rm M}^{\rm curr} E &=&
           \fr{8\alpha^4\psi(0)^2}{3mM}
           \lb 1-2\fr{\mu^2}{mM} \rb
           \ln^2\alpha, \label{Mcurr}  \\
\delta_{\rm M}^{\rm ret} E &=&
           - \fr{16\alpha^4\psi(0)^2}{15mM}
           \lb 1-\fr{5}{2}\fr{\mu^2}{mM} \rb
             \ln^2\alpha, \label{Mret} \\
\delta_{\rm M}^{\rm magn} E &=&
           -\fr{8\alpha^4\psi(0)^2\mu^2}{3m^2M^2}
           \ln^2\alpha.     \label{Mmagn}
\ea

4. The next source of the double logarithmic corrections 
is the double magnetic exchange with two seagull vertices.
The corresponding potential reads
\be\label{vSS}
V_{\rm SS} = - \fr{\alpha^2}{mM}
            \int \fr{d^3 k}{(2\pi)^3}
            \fr{4\pi}{2k} \fr{4\pi}{2k'}
            e^{i\q\r_p}
            \fr{(\vec{e}\vec{e}')^2}{
            k+k'+H-E}
            e^{-i\q\r_e}.
\ee
Here $\k'=\q-\k$, $e_i e_j=\delta_{ij}-k_i k_j/k^2$
and similarly for $\vec{e}'$. It is sufficient
to consider  the leading
nonrelativistic approximation for the seagull
vertex. From Eq.(\ref {vSS}) one sees, that 
the logarithmic contribution comes from the region of momenta 
where $|\q| \ll |\k| \ll \mu$. Neglecting $H-E$ in comparison with $k$
and integrating over $k$ from $q$ to
$\mu$  one obtains:
\be\label{vSSLO}
V_{\rm SS}^{\rm LO} = \fr{2\alpha^2}{mM}
            \ln \fr{q}{\mu}.
\ee

We consider now   relativistic corrections to the operator 
in Eq.(\ref{vSSLO}). In this case there are other sources of  
corrections, as compared to two cases considered above. The 
following corrections should be considered -- relativistic
corrections to the seagull vertex, an appearance of the 
magnetic-magnetic-Coulomb vertex, an expansion of the ``heavy'' 
intermediate energy denominators in powers of $|\p|/m,k/m,k'/m$
and the usual\footnote{By ``usual'' we mean here such retardation
effects, that do not resolve the point-like seagull vertex.}
retardation effects. We find that none of these effects
produces the double logarithmic correction. 

For example, relativistic corrections to the seagull vertex 
are polynomial in the momenta $p,~k$ and $q$. On the other hand, 
the structure of the denominator in Eq.(\ref{vSS}) does not change.
In this case the logarithmic contribution comes only from 
the region of large $|\k|$, so that the resulting operator is
of the form of Eq.(\ref{vSSLO}) times a polynomial in external
momenta. Such operators are too singular to produce a double
logarithmic contribution. 

Hence we conclude that the ${\cal O}(m\alpha^7\ln^2 \alpha)$
corrections to the expectation value of the operator 
Eq.(\ref{vSSLO}) appear only due to 
relativistic corrections to the wave functions. 
As we have mentioned
already, these corrections are considered at the
end of this Letter.

5. We now turn to logarithmic operators which
arise at the ${\cal O}(m\alpha^6)$ order.
The ${\cal O}(\alpha)$ correction
to such operators can be caused only by the effect of
retardation. There are two effects of this sort. 
Let us denote by $k$ the momentum
of the photon in the intermediate state. Then,
for the ``light" $e^+e^-\gamma$ intermediate state
the energy denominators are expanded  in $(H-E)/k$,
whereas for the ``heavy" $e^+ e^- e^+ e^- \gamma $ intermediate
state the expansion parameter is $k/m$.
In contrast to these, all other relativistic corrections
are of relative $\alpha^2$ order.
Retardation corrections clearly require that the corresponding
effective operators are induced by an exchange of at least
one magnetic photon. Inspecting such operators in \cite{CMY}, 
we observe that only two of them, the ``one-loop" operator and the
operator induced by a double magnetic exchange with one
seagull vertex, can give rise to 
${\cal O}(m\alpha^7\ln^2 \alpha)$ corrections
on  one hand, and were not considered yet on the other.

Retardation correction to the one-loop operator 
cannot produce the double logarithmic contribution.  
Indeed, both effective
vertices entering the one-loop operator, are
$\p$-independent, so that their commutators with $H$
produce only positive powers of momenta. One can easily
check that resulting operators are too singular to
produce two logarithms.

For the same reason considering the double magnetic
exchange with one seagull vertex we keep orbital
currents only. The sum of three diagrams (see Fig.2) then gives:
\ba
V^{\rm 2ex}_{\rm S} &=& \fr{\alpha^2}{m}\int \fr{d^3k}{(2\pi)^3}
          \fr{4\pi}{2k'} \fr{4\pi}{2k}
          \left\{
          v_2 \fr{1}{k'+k+H-E} v_1 \fr{1}{k+H-E} v_3
          \right. \non \\
        && \left.
        + v_1 \fr{1}{k'+H-E} v_2 \fr{1}{k+H-E} v_3
        + v_1 \fr{1}{k'+H-E} v_3 \fr{1}{k'+k+H-E} v_2
          \right\},  \label{seag}
\ea
where
\be
v_1 = \fr{\p'\vec{e}'}{M}e^{i\k'\r'_p}, \qquad
v_2 = \vec{e}'\vec{e}e^{i\q\r_e}, \qquad
v_3 = \fr{\p\vec{e}}{M}e^{i\k\r_p}.
\ee
The contribution of three diagrams with the seagull
vertex on the second particle line is obtained
from Eq.(\ref{seag}) by the substitution $m \leftrightarrow
M$. An account of the retardation to 
first order gives the operator:
\ba
V^{\rm 2ex}_{\rm S} &\to& - \fr{\alpha^2}{m}\int
          \fr{d^3k}{(2\pi)^3}
          \fr{4\pi}{2k'} \fr{4\pi}{2k}
          \left\{
          v_2 \fr{1}{k'+k} v_1 \fr{1}{k^2}[H, v_3]
        + v_1 \fr{1}{k'} v_2 \fr{1}{k^2} [H,v_3]
          \right. \non \\
       && \left.
        + v_1 \fr{1}{k'} v_3 \fr{1}{(k'+k)^2} [H,v_2]
          \right\} + {\rm h.c.}.
\ea
For $k \ll q$ this operator reduces to
\be
V^{\rm 2ex}_{\rm S} = \fr{\alpha^3}{3mM^2}
        \int \fr{d^3k}{(2\pi)^3}
        \fr{4\pi}{k^3} \fr{4\pi}{q^2}
        \p_q [\p_q,C(r)]
        + {\rm h.c.}
        + (m \leftrightarrow M).
\ee
Double logarithmic correction to the energy arises if
one integrates over $k$ from
$q^2/\mu$ to $q$, and then over $q$ from $\mu\alpha$ to
$\mu$. The result is
\be
\delta^{\rm 2ex}_{\rm S} E = - \fr{8\alpha^4\psi(0)^2}{3mM}
                     \ln^2\alpha. \label{2S}
\ee

The last ``irreducible" correction arises due to the single-seagull 
diagrams with one of the magnetic quanta absorbed by the same charged 
particle (see Fig.3). In contrast to the previous calculation, 
there is a double logarithmic contribution in the diagrams Fig.3a,b
coming from the region of $\q \ll k \ll \mu$. In the sum of Fig.3a and
Fig.3b the contribution of this region cancels out. 
The result is
:
\be
\delta^{\rm 1ex}_{\rm S} E = - \fr{8\alpha^4\psi(0)^2}{3mM}
                     \ln^2\alpha. \label{1S}
\ee

Eq.(\ref{1S})  completes the analysis of the ``irreducible''
${\cal O}(m\alpha^7\ln^2\alpha)$ corrections.

6. The last source of the ${\cal O}(m\alpha^7 \ln^2 \alpha)$
corrections is related to the wave function modification
by  relativistic effects:
\be
\delta_{\psi} E = \la VGU + UGV \ra.
\ee
Here $G$ is the reduced nonrelativistic
Green function, $U$ is the Breit Hamiltonian
projected on $S$-states:
\be
U(\p,\r) = -\fr{1-3\fr{\mu^2}{mM}}{2\mu}
            \lb \fr{p^2}{2\mu} \rb^2
           + \fr{\mu}{mM}
            \left\{ \fr{p^2}{2\mu}, C(r) \right\}
           + \fr{\pi\alpha}{2\mu^2}
           \lb 1 + 2\fr{\mu^2}{mM}
           \left[ 1 + \fr{2}{3}\sS \right] \rb
           \delta(\r),
\ee
and the operator $V$ is the sum of the lowest-order logarithmic
operators Eqs.(\ref{4},\ref{15},\ref{vSSLO}):
\be
V(\p,\r) = -\fr{2\alpha}{3\pi\mu^2}
        \ln \fr{p^2}{\mu^2}
        \left\{
        \lb 1 - \fr{\mu^2}{mM} \rb
        \p (H-E) \p
        - \fr{3\pi\alpha}{2}\fr{\mu^2}{mM}
        \right\}.
\ee
The sum of the first two terms in $U(\p,\r)$ can be represented
as a linear combination of the operators $H^2$,
$\{H,C(r)\}$, and $C(r)^2$ (see \cite{CMY}).
The last two terms induce the following double
logarithmic corrections:
\ba
\delta_{\psi,1} E &=& \fr{1-\fr{\mu^2}{mM}}{\mu}
                \la \{H,C(r)\} G V \ra
                \to - \fr{1-\fr{\mu^2}{mM}}{\mu}
                \la C(r) V \ra
                \non \\
         &\to&  -\fr{2\alpha}{3\pi\mu^3}
                \lb 1-\fr{\mu^2}{mM} \rb^2
                \la \ln \fr{p^2}{\mu^2}
                C(r) \p [\p,C(r)] \ra
                \non \\
         &\to&  \fr{16\alpha^4\psi(0)^2}{3\mu^2}
                \lb 1-\fr{\mu^2}{mM} \rb^2
                \int \fr{d^3 p}{(2\pi)^3}
                \fr{\pi^2}{p^3} \ln \fr{p^2}{\mu^2}
                \non \\
         &\to&  -\fr{8\alpha^4\psi(0)^2}{3\mu^2}
                \lb 1-\fr{\mu^2}{mM} \rb^2
                \ln^2 \alpha;
\ea
\ba
\delta_{\psi,2} E &=& - \fr{1+\fr{\mu^2}{mM}}{\mu}
                \la C(r)^2 G V \ra
                \non \\
         &\to&  -\fr{1+\fr{\mu^2}{mM}}{\mu^3}
                \la \ln \fr{p^2}{\mu^2}
                C(r)^2 G \left\{
                \lb 1-\fr{\mu^2}{mM} \rb
                \p [\p,C(r)]
                + \fr{3\pi\alpha}{2}\fr{\mu^2}{mM}
                \right\} \ra
                \non \\
         &\to&  \fr{1+\fr{\mu^2}{mM}}{\mu^3}
                \la \ln \fr{p^2}{\mu^2}
                C(r)^2 G_0 \left\{
                4\pi\alpha \lb 1-\fr{\mu^2}{mM} \rb
                -\fr{3\pi\alpha}{2}\fr{\mu^2}{mM}
                \right\} \ra
                \non \\
         &\to&  \fr{16\alpha^4\psi(0)^2}{3\mu^2}
                \lb 1-\fr{3}{8}\fr{\mu^2}{mM}
                -\fr{11}{8}\fr{\mu^4}{m^2M^2}\rb
                \ln^2 \alpha.
\ea

In a similar manner we consider the contribution of the
$\delta(\r)$ operator from the Breit Hamiltonian.
We obtain:
\ba
\delta_{\psi,3} E &\to &
           \fr{\pi\alpha}{\mu^2}
           \lb 1 + 2\fr{\mu^2}{mM}
           \left[ 1 + \fr{2}{3}\sS \right] \rb
               \la \delta(\r) G V \ra
               \non \\
      &\to& - \fr{4\alpha^4\psi(0)^2}{3\mu^2}
            \lb 1 + 2\fr{\mu^2}{mM}
           \left[ 1 + \fr{2}{3}\sS \right] \rb
           \lb 1 -\fr{7}{4}\fr{\mu^2}{mM} \rb
           \ln^2 \alpha. \label{psi2}
\ea

7. In order to check that we have accounted for  all sources of 
the ${\cal O}(m\alpha^7\ln^2\alpha)$ corrections, 
it is useful to consider  the problem from a more formal point of
view. 
As we mentioned above, logarithms of $\alpha$ arise because of the 
hierarchy of scales in QED bound states calculations. 
In the framework of NRQED such logarithms may be 
detected by analyzing the on-shell scattering amplitude 
of two particles  and identifying such 
contributions to  this amplitude that have zero index in the 
nonrelativistic region. 
Therefore, to get the ${\cal O}(m\alpha^7\ln^2\alpha)$ 
corrections to  energy levels, we have to consider 
the ${\cal O}(\alpha^4)$ scattering amplitude and
find all zero index graphs.  

To do so, we proceed in the following way: i) divide all graphs 
into classes according to the number of magnetic photons 
(e.g., class 0 contains all graphs with four Coulomb exchanges, 
class 1 contains all graphs with three Coulomb and one magnetic
photon, and so on); ii) in each class, we consider only such graphs 
that in the leading nonrelativistic approximation have negative 
or zero indices; iii) the negative indices are shifted to zero 
by accounting for  retardation corrections as well as all possible 
relativistic corrections to the elements of a given graph. 

Consider first  the class 0. In the leading non-relativistic approximation
the only contribution comes from the four-Coulomb ladder graph with 
both electron and positron  staying in positive-energy intermediate 
states. This graph has the index 
$3\times 3 - 3\times 2 - 4\times 2 =- 5$, where the first term comes 
from three-loop integration volume, the second one from three 
energy denominators, corresponding to $e^+e^-$ intermediate states and 
the last term arises from the four Coulomb propagators. To shift 
this index to zero, we have to consider relativistic corrections 
with the fifth net power of momenta. However, all relativistic 
effects in pure Coulomb graphs bring in corrections which scale
like ${\cal O}(v^2)$, i.e. they increase the index by $2$. 
Therefore, we conclude that the class 0 is free from potentially 
logarithmic graphs.

Turning to other classes, we notice that one can significantly 
decrease the number of graphs by considering only such where all 
magnetic and seagull vertices belong to 
one and the same fermion line. 
It is easy to see that the equal-time transfer of a 
vertex from one fermion line to the other does not change the index of 
the graph. Hence, for the purpose of  index counting we can consider 
the simplified problem of the ${\cal O}(\alpha^4)$ scattering of one 
particle at the external Coulomb field. After identifying
potentially logarithmic graphs with magnetic and seagull vertices 
on one fermion line one must consider all possible equal-time
transfers of those vertices between two fermion lines and 
all possible relativistic corrections.

To make further discussion more transparent, we introduce the
following notations. The Coulomb, magnetic, and seagull vertices 
are denoted by C, M, and S, respectively. 
For the class 1, in the leading nonrelativistic approximation we have 
the following graphs: MMCCC, CMMCC, CCMMC, and CCCMM\footnote{The
notations are such, that e.g. MMCCC represents a graph were
first the magnetic photon is emitted, then it is absorbed and
then three Coulomb exchanges occur.}.
Their index, -3, is the sum of 9 from the integration volume, 
2 from two Pauli currents in magnetic vertices, --6 from three 
Coulomb propagators, --6 from three energy denominators related to 
intermediate states without photon, --1 from the $e^+e^-\gamma$ 
intermediate state, and --1 from the  normalization factor of the 
propagator of the magnetic photon. To increase this index to
zero it is necessary to account for the effects of retardation. 
Since each order of  perturbation theory for the retardation
increases the index by 1, we have to consider either the third order
retardation effects or the first order retardation with 
additional ${\cal O}(v^2)$ relativistic corrections. 
Note that an account of the retardation includes not only an 
expansion of energy denominators $(k+\Delta E)^{-1}$, but also
permutations of the Coulomb and magnetic vertices, e.g., 
MMCCC$\to$MCMCC.  By inspection we find that all these effects 
were considered  in our previous analysis of single-magnetic 
contributions.

An analysis of the class 2 graphs is slightly more involved because of
an appearance of the seagull vertex. The lowest index, --2, arises in 
the leading nonrelativistic approximation for the graphs with two 
seagull vertices, CCSS, CSSC, and SSCC. To compensate this index we
have to account for the next-to-next-to-leading order retardation 
or some ${\cal O}(v^2)$ relativistic correction. The index --1 
graphs include SMM, MSM, MMS, 
and ${\rm M}_i{\rm M}_i{\rm M}_j{\rm M}_j$
with two additional Coulomb vertices, none of which 
is located either between S and M or ${\rm M}_i$ and ${\rm M}_i$.
The next-to-leading order retardation shifts 
the index of these graphs to zero. Finally, there are index 0 graphs, 
${\rm M}_i{\rm M}_j{\rm M}_j{\rm M}_i$ and 
${\rm M}_j{\rm M}_i{\rm M}_j{\rm M}_i$ with two additional 
Coulomb photons,  each  located either to the left or 
to the right of all M-vertices. One can easily  find the one-to-one 
correspondence between any  of these graphs and some parts of the 
double-magnetic contributions  considered in our previous analysis. 

In the class 3 there are only index 0 graphs, SSMM and MMSS with one C 
outside pairs SS and MM. After equal-time transfer of one seagull 
and one magnetic vertex to the second fermion line, we get the graphic
representation for the correction to the average value of the operator 
Eq.(\ref{vSSLO}), induced by  the magnetic part of the 
Breit Hamiltonian.

Finally, an inspection of the class 4 graphs shows
that all of them have positive indices.

8. It remains to sum up all ${\cal O}(m\alpha^7\ln^2\alpha)$
contributions. We obtain:
\ba
\delta E_{\rm aver} &=&
            - \fr{\alpha^4\psi(0)^2}{\mu^2}
            \lb 4 - \fr{149}{15}\fr{\mu^2}{mM} \rb
            \ln^2 \alpha,
            \label{Eaver}\\
\delta E_{\rm hfs} &=&
            - \fr{64\alpha^4\psi(0)^2}{9mM}
            \lb 1 -\fr{7}{4}\fr{\mu^2}{mM} \rb
            \ln^2 \alpha.
\ea
In the limiting case $m/M \to 0$, the old
result for the hydrogen \cite{LFY} is reproduced.
Considering pure recoil corrections, one finds:
\be
\delta E_{\rm aver}^{\rm rec} =
            - \fr{11\alpha^4\psi(0)^2}{15mM}
            \ln^2 \alpha, 
\qquad   \delta E_{\rm hfs}^{\rm rec} =
            \fr{16\alpha^4\psi(0)^2
            \mu^2}{3m^2M^2}
            \ln^2 \alpha.
\ee
It is interesting to note, that the ${\cal O}(\mu^2/(mM)^2)$ terms  
mutually canceled not only in Eq.(\ref{Eaver}), but also in  
individual contributions, i.e. those induced by the self energy, the 
single magnetic exchange and the double magnetic
 exchange with one seagull vertex.

For the positronium, it is necessary to add the
contribution caused by the
virtual annihilation which is easily extracted from
Eq. (\ref{psi2}). One obtains:
\be
\delta_{\rm ann} E = - 3 \fr{ 3+\sS }{4}
            \fr{\alpha^4\psi(0)^2}{m^2}
            \ln^2 \alpha.
\ee

Using  $\psi(0)^2=\delta_{l0}
m^3 \alpha^3/(8\pi n^3)$, we arrive at the final
result for the ${\cal O}(m\alpha^7\ln^2\alpha)$
contribution to the positronium energy levels:
\be
\delta E = - \lb \fr{499}{15}+7\sS \rb
\fr{m\alpha^7 \ln^2 \alpha}{32\pi n^3} \delta_{l0}.
\ee
Its spin-dependent part is in agreement with the result
of Ref.\cite{DL}.

Numerically, the ${\cal O}(m\alpha^7\ln^2\alpha)$ contribution
to the triplet energy levels equals to
\be
\delta E \lb n^3 S_1 \rb = - \fr{1.3}{n^3} \; {\rm MHz}.
\ee
The correction to the difference $E(2^3 S_1)-E(1^3 S_1)$
amounts therefore to 1.16 MHz.
For the singlet states, the ${\cal O}(m\alpha^7 \ln^2\alpha)$
correction gives:
\be
\delta E \lb n^1 S_0 \rb = - \fr{0.40}{n^3} \; 
{\rm MHz}.
\ee
We then find that the ${\cal O}(m\alpha^7 \ln^2\alpha)$ correction 
to the positronium ground state hyperfine splitting
 amounts to $-0.9$ MHz.

We  conclude, that the ${\cal O}(m\alpha^7 \ln^2 \alpha)$
corrections turn out to be of the order of $1$ MHz and hence
somewhat enhanced, as compared to the naive estimate of the magnitude 
of the $m\alpha^7$ effects. We note, that at $O(m\alpha^6)$ the 
remaining, non-logarithmic corrections, contributed approximately
one half of the leading logarithmic ones. Extrapolating this
situation to order ${\cal O}(m\alpha^7)$, we conclude that 
the current theoretical uncertainty in the predictions 
for the positronium energy levels \cite{KP,CMY} should be 
approximately $0.5$ MHz.

9. Acknowledgments. 
We are grateful to K. Pachucki for communicating to us
some of the results prior to publication and for 
comments on the manuscript.  
We are grateful to A. Czarnecki for
providing the figures.  The research of K. M. was 
supported in part by BMBF under grant number BMBF-057KA92P, 
and by Graduiertenkolleg ``Teilchenphysik'' at the University 
of Karlsruhe. A.Y. was partially supported by the Russian
Foundation for Basic Research under grant number 98-02-17913.

{\em Note added:} 
We are indebted to K. Pachucki and S. Karshenboim for pointing out 
an error in Eq.(\ref{1S}) in the previous version of this Letter. 
After correcting this error, our result for the 
${\cal O}(m\alpha^7 \log^2 \alpha )$ corrections to positronium 
energy levels agrees with that of Ref. \cite{KP1}.

\begin{figure}[h]
\hspace*{-5mm}
\begin{minipage}{10.cm}
\[
\mbox{
\hspace*{40mm}
\begin{tabular}{cc}
\psfig{figure=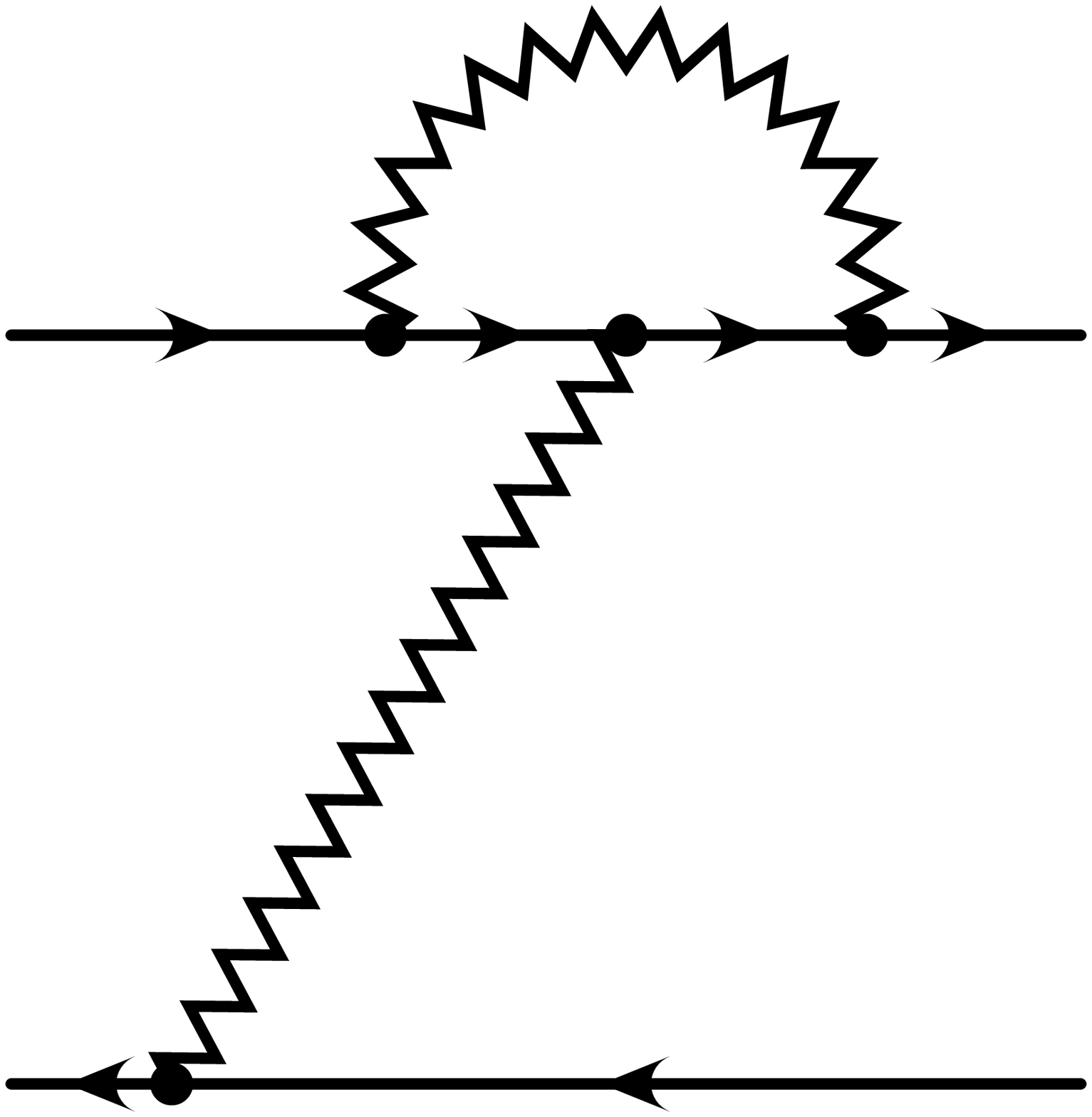,width=25mm}
&\hspace*{10mm}
\psfig{figure=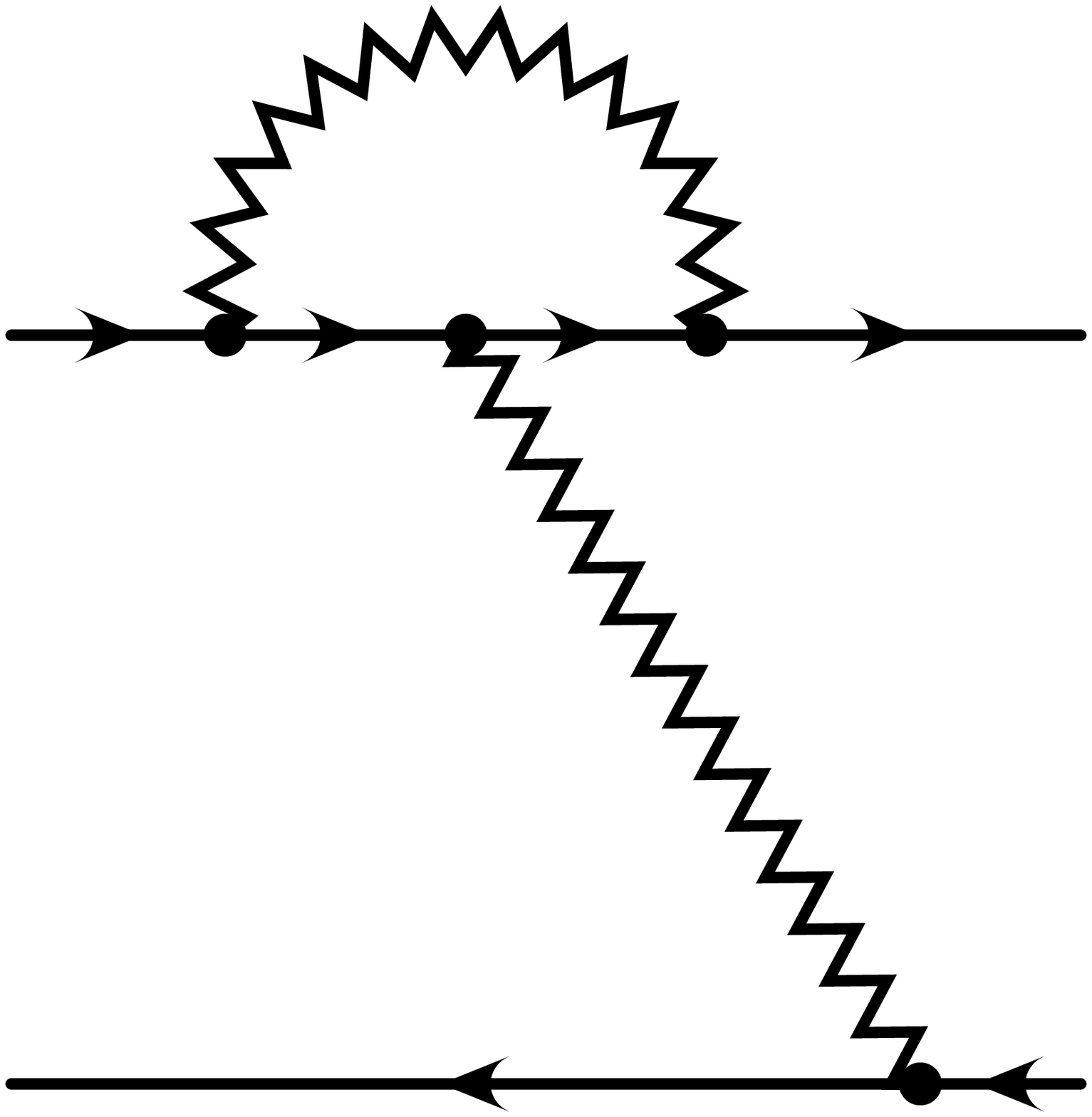,width=25mm}
\\[2mm]
(a) &\hspace*{10mm} (b)
\\[5mm]
\psfig{figure=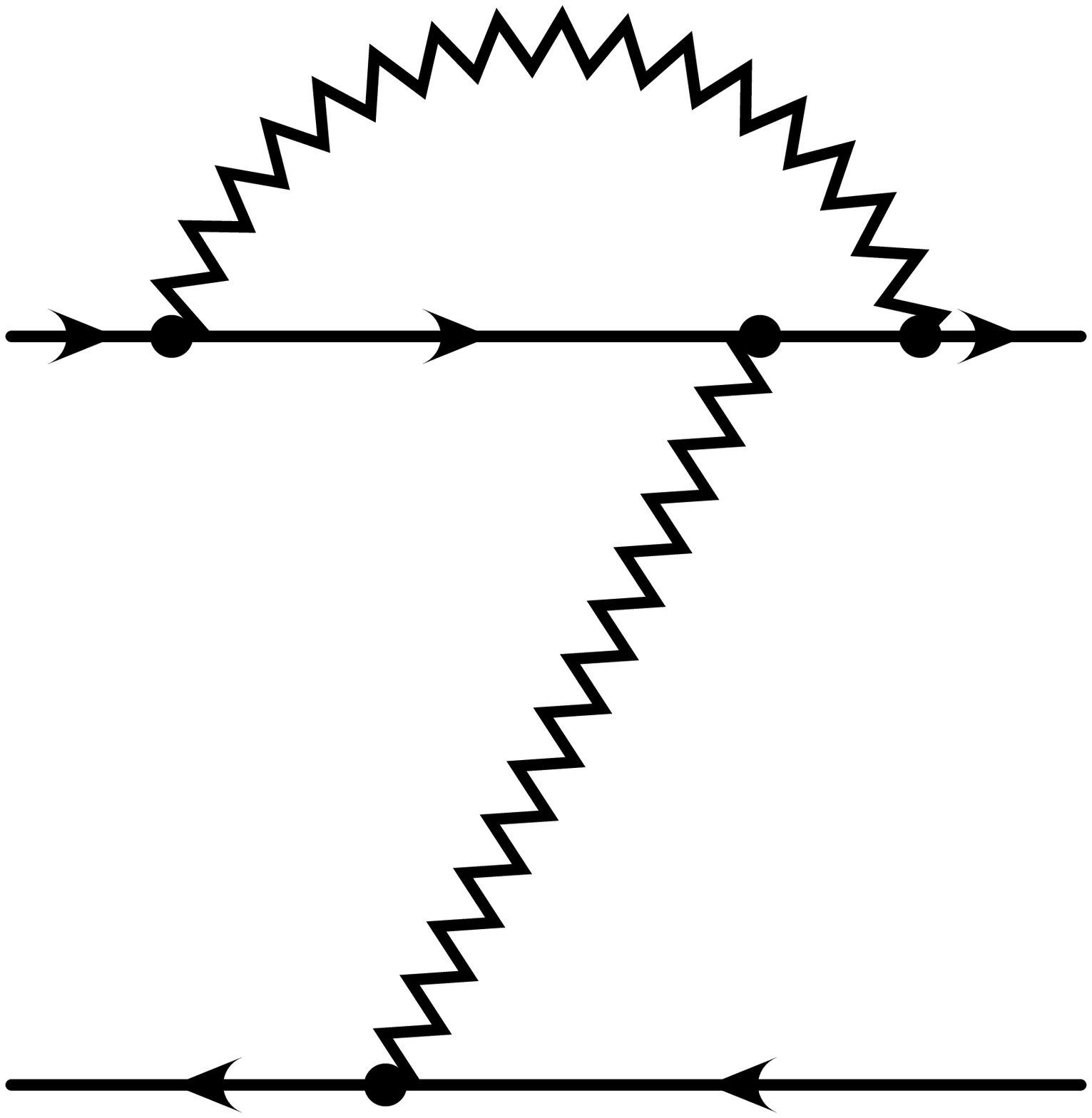,width=25mm}
&\hspace*{10mm}
\psfig{figure=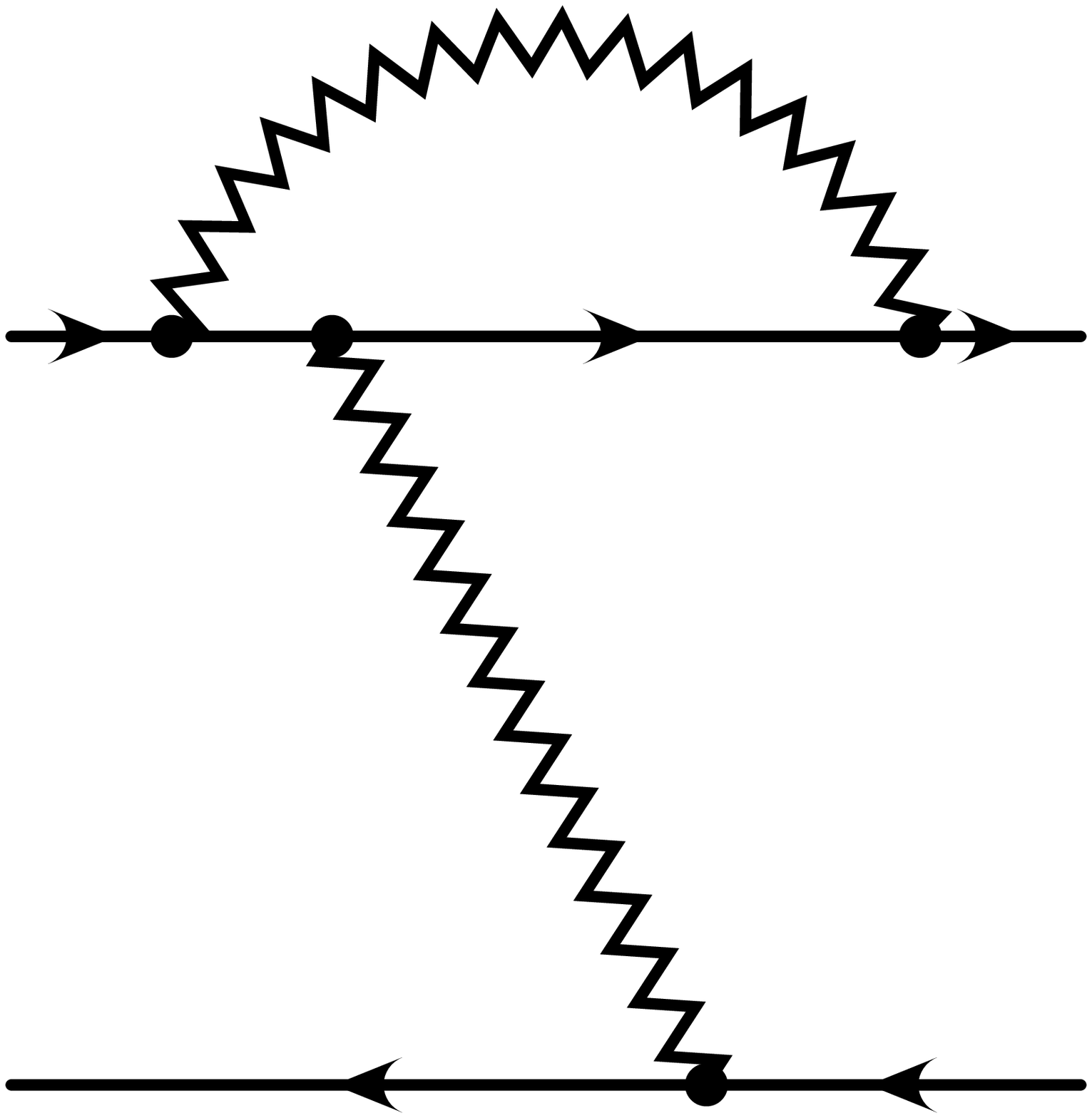,width=25mm}
\\[2mm]
(c) &\hspace*{10mm} (d)
\end{tabular}}
\]
\end{minipage}
\caption{Corrections to self-energy operator
Eq.(\ref{4}) due to additional magnetic exchange.}
\label{fig:ris1}
\end{figure}

\begin{figure}[h]
\hspace*{-5mm}
\begin{minipage}{10.cm}
\[
\mbox{
\hspace*{20mm}
\begin{tabular}{ccc}
\psfig{figure=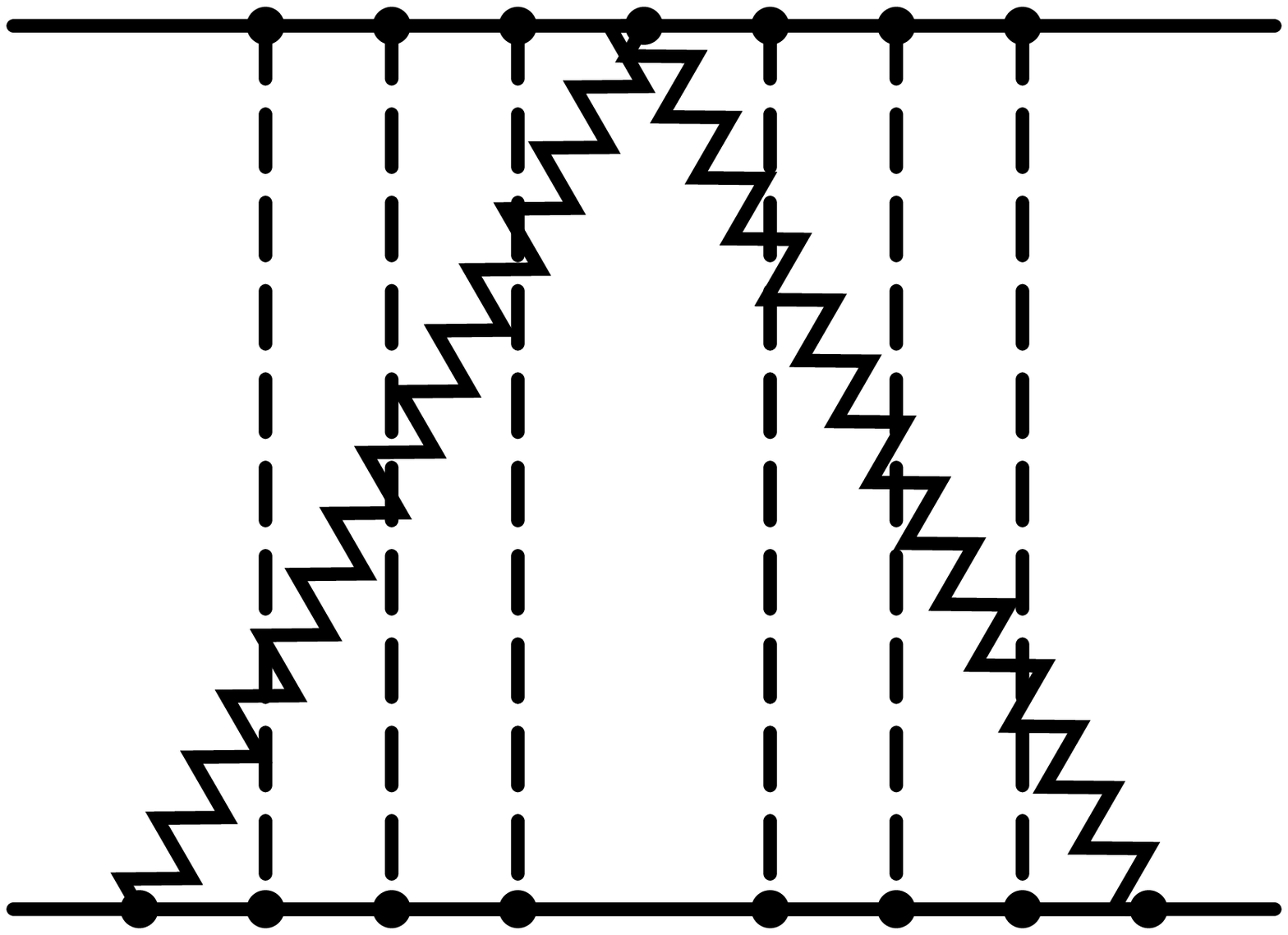,width=30mm}
&\hspace*{10mm}
\psfig{figure=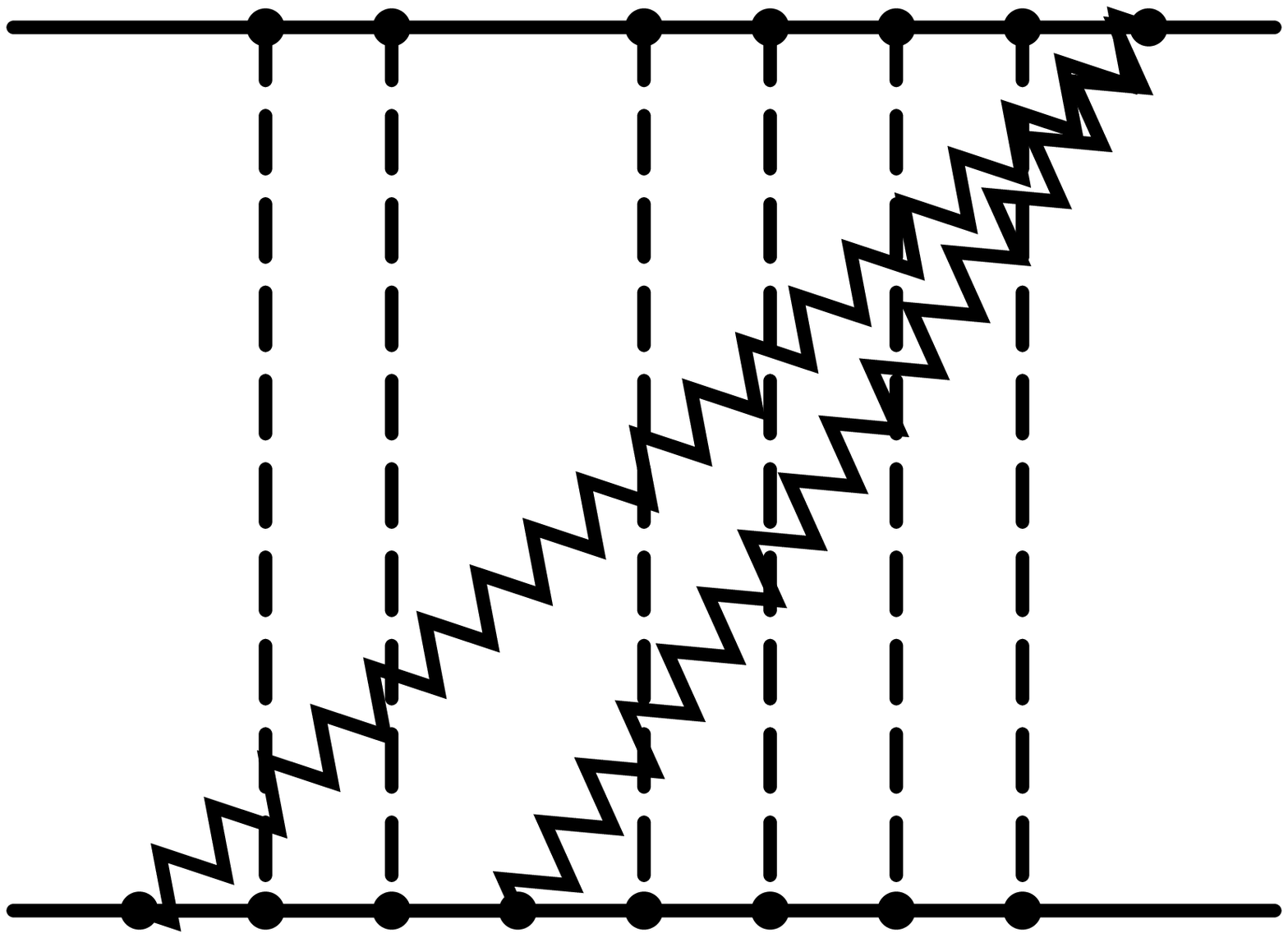,width=30mm}
&\hspace*{10mm}
\psfig{figure=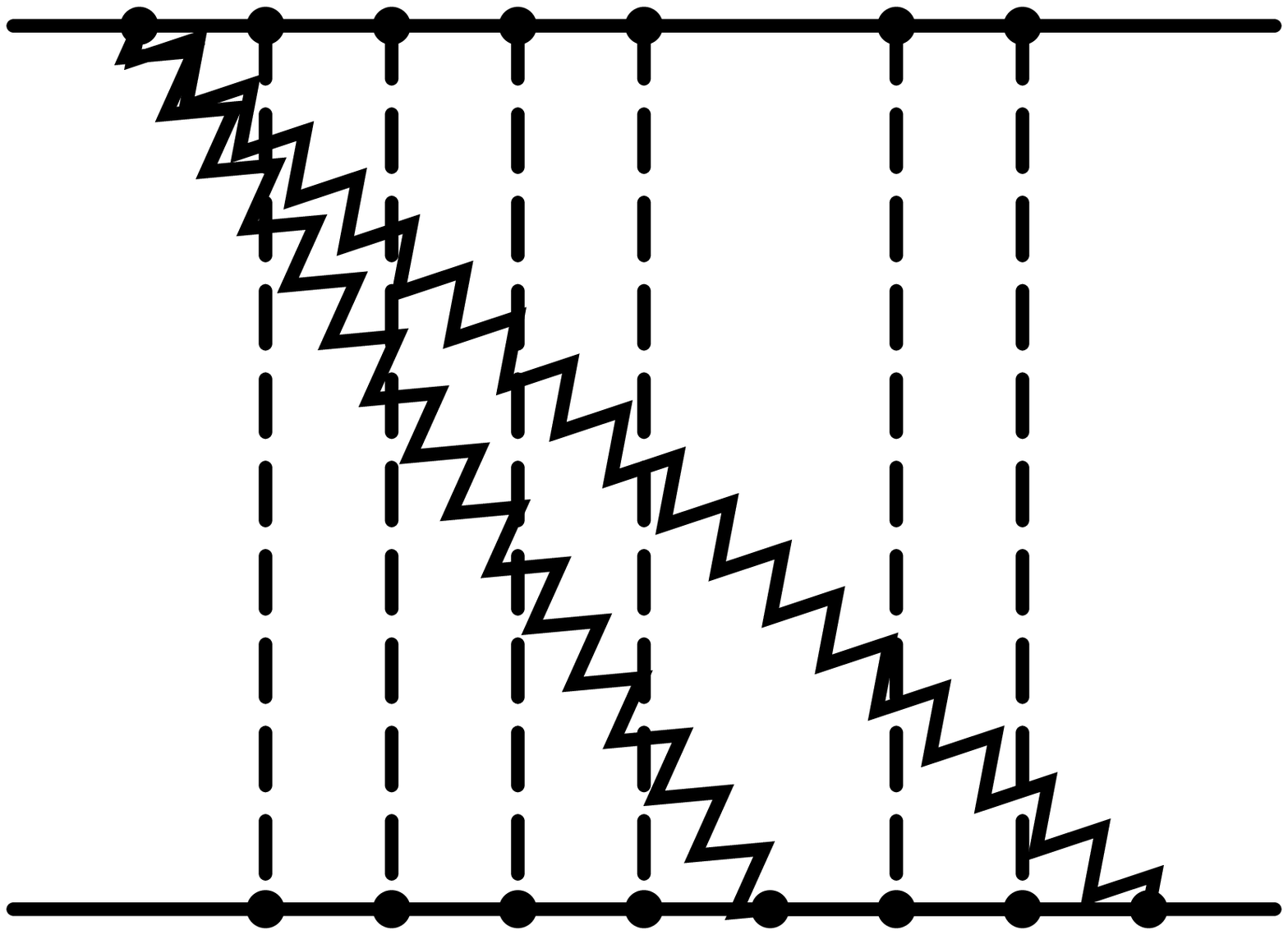,width=30mm}
\\[4mm]
(a) &\hspace*{10mm} (b)&\hspace*{10mm} (c)
\end{tabular}}
\]
\end{minipage}
\caption{Double magnetic exchange with one-seagull vertex.
The dashed lines, representing Coulomb exchanges, 
show that the exact Green function for the system positronium 
plus photon(s) should be used in the calculation.}
\label{fig:ris2}
\end{figure}

\begin{figure}[h]
\hspace*{-5mm}
\begin{minipage}{10.cm}
\[
\mbox{
\hspace*{20mm}
\begin{tabular}{ccc}
\psfig{figure=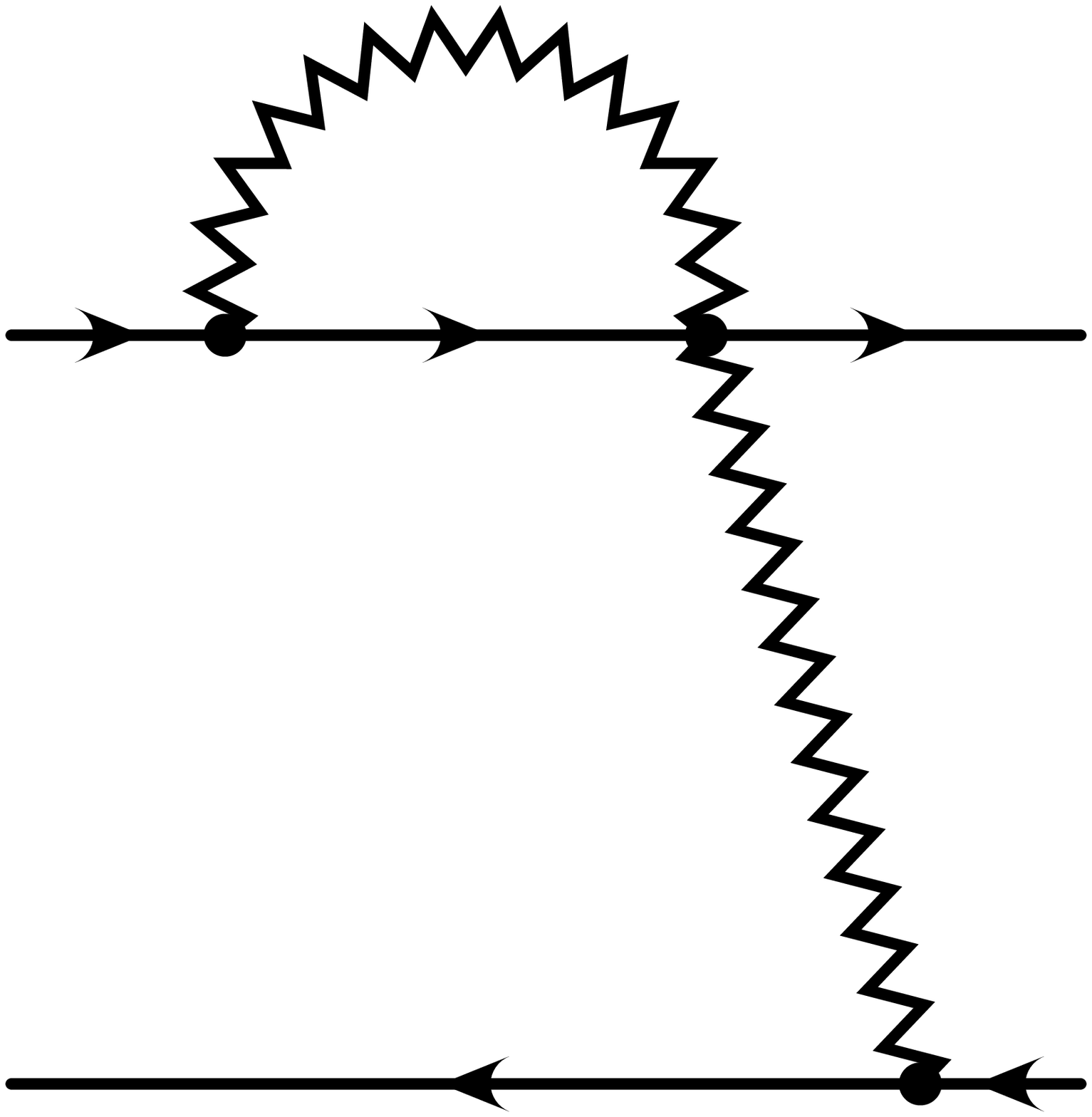,width=25mm}
&\hspace*{10mm}
\psfig{figure=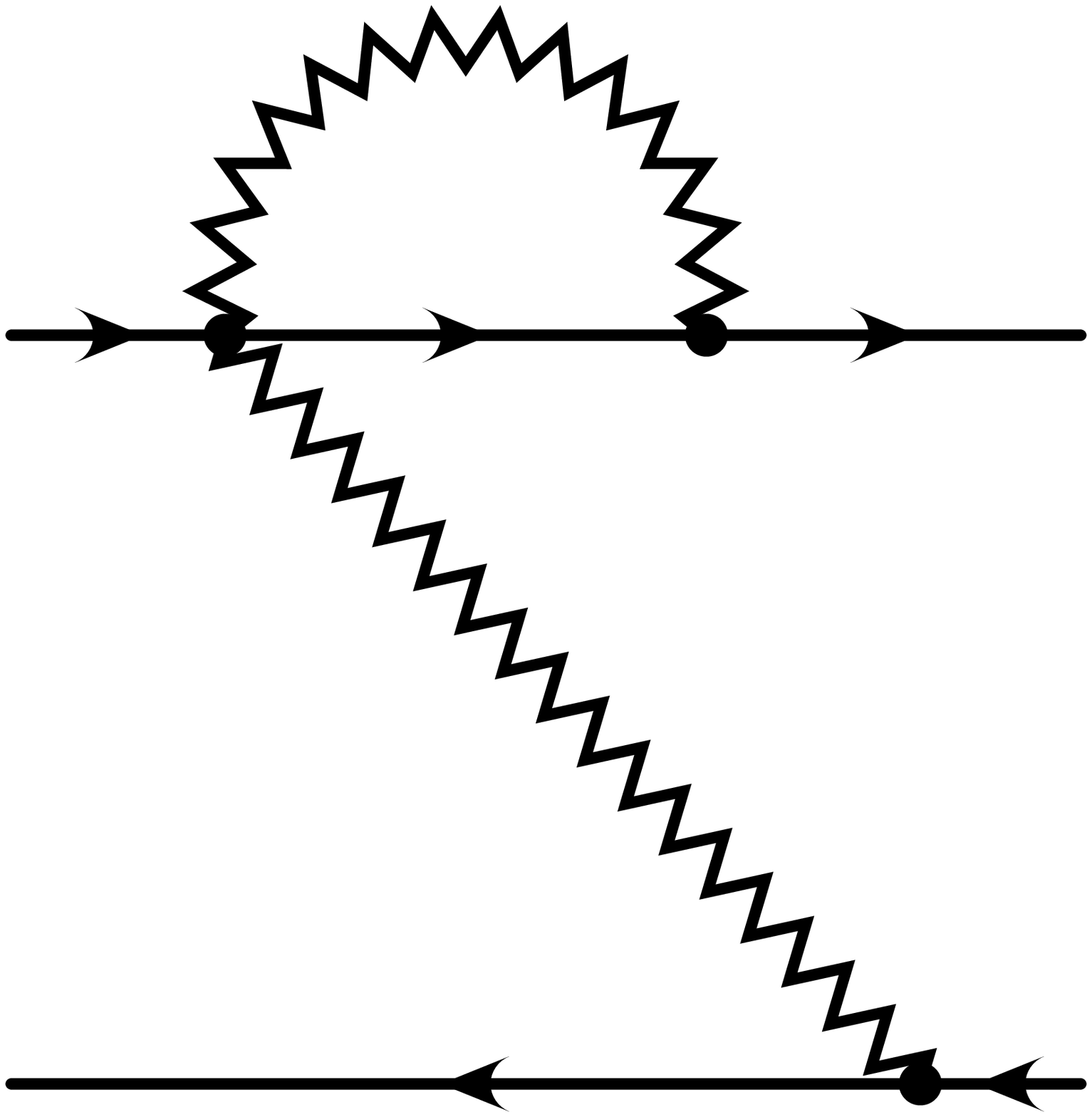,width=25mm}
&\hspace*{10mm}
\psfig{figure=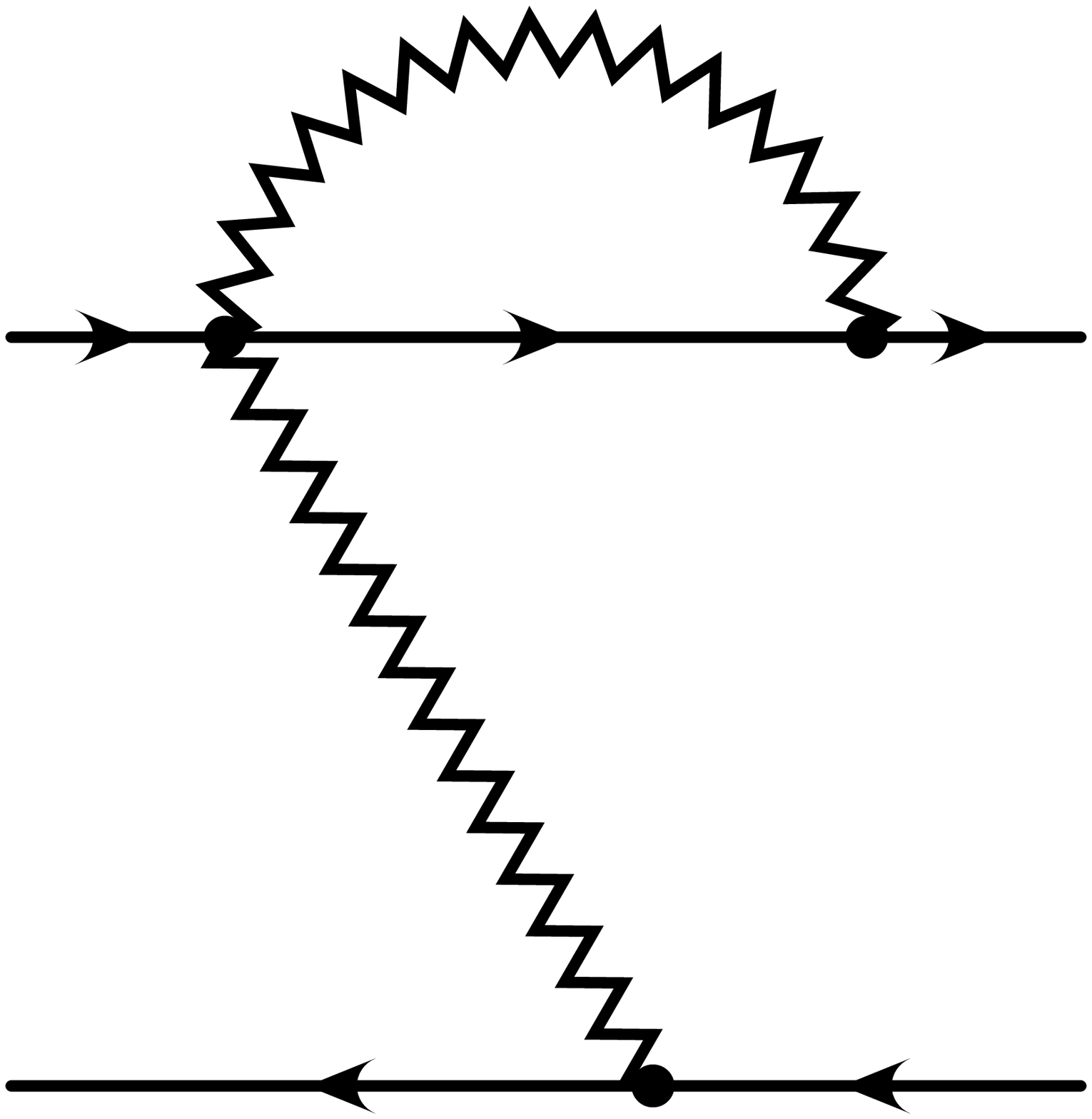,width=25mm}
\\[4mm]
(a) &\hspace*{10mm} (b)&\hspace*{10mm} (c)
\end{tabular}}
\]
\end{minipage}
\caption{Seagull corrections to single magnetic exchange.}
\label{fig:ris3}
\end{figure}

\end{document}